\documentclass[12pt]{article}
 
\usepackage{amssymb}
\usepackage{amsmath}
\usepackage{amsfonts}
\usepackage{amscd}
\usepackage{graphicx}
\usepackage{amsthm}
\usepackage{setspace}
\usepackage{epsfig} 
\usepackage{bm}
\usepackage{amssymb}
\usepackage{hyperref}
\usepackage[authoryear,round]{natbib}
\usepackage{booktabs}
\usepackage{threeparttable}
\usepackage{float}
\usepackage{subfig}

\newtheorem{theorem}{{\bf \sc Theorem}}
\newtheorem{proposition}{{\bf \sc Proposition}}
\newtheorem{lemma}{{\bf \sc Lemma}}
\newtheorem{corollary}{{\bf \sc Corollary}}

\theoremstyle{definition}

\newcommand{\be}{\begin{equation}}
\newcommand{\ee}{\end{equation}}
\newcommand{\bes}{\begin{equation*}}
\newcommand{\ees}{\end{equation*}}

\def\eproof{\hfill \hbox{\hskip3pt\vrule width4pt height8pt depth1.5pt}}

\newcommand{\Ex}{\mathbb E}

\DeclareMathOperator{\Var}{Var}

\allowdisplaybreaks 

\begin{document}

\title{\textbf{Strategic Random Networks} \\ Why Social Networking Technology Matters\thanks{We thank Jeremy Bulow, Matthew Elliott, Alex Frankel, Matthew O. Jackson, Alex Hirsch, Ruth Kricheli, Antoine Lallour, Carlos Lever, James Liang and seminar participants at Stanford for helpful conversations.}}
\date{March 22, 2010}
\author{Benjamin Golub\thanks{\mbox{Graduate School of Business, Stanford University.
Email: bgolub@stanford.edu,}\hfill \mbox{http://www.stanford.edu/$\sim$bgolub/}} \and %
 \and Yair Livne\thanks{\mbox{Graduate School of Business, Stanford University.
Email: ylivne@stanford.edu}}}\maketitle

\begin{abstract} This paper develops strategic foundations for an important statistical model of random networks with heterogeneous expected degrees. Based on this, we show how social networking services that subtly alter the costs and indirect benefits of relationships can cause large changes in behavior and welfare. In the model, agents who value friends and friends of friends choose how much to socialize, which increases the probabilities of links but is costly. There is a sharp transition from fragmented, sparse equilibrium networks to connected, dense ones when the value of friends of friends crosses a cost-dependent threshold. This transition mitigates an extreme inefficiency.
\smallskip


\end{abstract}
\thispagestyle{empty}

\newpage
New Internet services that help people to track, maintain, and create social relationships have attracted hundreds of millions of users over the past decade. It has been claimed that these innovations --- websites such as Facebook, MySpace, LinkedIn, and Twitter --- fundamentally change the network formation dynamics of the groups that use them. For example, a 2007 \emph{New York Times} article stated that ``Facebook and other social networks like MySpace have transformed the social lives of teenagers in many ways, and that includes how they make the transition from high school to college'' \citep{NYTimesTransformed}. Why did this technology, which was an incremental advance in electronic communication, lead to such apparently large changes in social interaction? We develop a theory to address that puzzle, which explains how small changes in the costs and benefits of direct and indirect relationships can lead to large changes in equilibrium networks. 

The model that we develop for this purpose connects the economic theory of rational network formation with the random graphs that have become workhorses for modeling social networks in statistics, physics, and computer science --- namely those of \cite{chunglu}, built on the seminal work of \cite{ErdosRenyi}. While this powerful statistical model can fit a wide variety of empirically important networks with arbitrary degree distributions\footnote{The degree of an agent is the number of connections he has. Power law distributions, in which the proportion of nodes with degree $d$ is falls off as a power of $d$, have been observed in many applications \citep{NewmanSurvey}. As emphasized by Chung, Lu, and Vu \citeyearpar{ChungLuVu}, the Chung-Lu framework gives a tractable probabilistic model of such networks.} it lacks rational foundations and an understanding of how linking probabilities relate to more fundamental parameters, such as costs and benefits. These are important when the agents who populate the networks have some control over their interaction. Indeed, as emphasized by \cite{JacksonBook} and K\"{o}nig, Tessone, and Zenou \citeyearpar{formationZenou}, the random graphs literature has quite successfully addressed ``how'' links seem to form, but has less to say about ``why'' and about where the parameters in the models come from. To provide rational foundations for these models, and to address our motivating puzzle, we focus on a simple environment. Agents, such as an entering cohort of students at a business school, meet each other for the first time and socialize now to create relationships that may be realized in the future. Those realizations are uncertain, but their probabilities increase with investment, as agents socialize more within the group. Agents value friends and also friends of friends. They trade off the expected direct and indirect benefits of links against the costs of socializing, which are agent-specific private information. This model is flexible enough to produce networks with arbitrary degree distributions and clustering, and it yields sharp predictions about the distributions of important network statistics such as connectedness, diameter, and density as functions of the key economic parameters. As a result, it suitable for structural modeling. 

This approach goes beyond giving economic interpretations to the parameters of an important statistical model.  The modeling reveals that there are surprising and highly nonlinear relationships between the economic fundamentals and the properties of the realized networks due to the ways agents best-respond to each other. These results are used to give one explanation of the Facebook puzzle.

\begin{table}
\begin{center}
\begin{threeparttable}
\begin{tabular}{@{\extracolsep{.1in}}p{2.1in}ll} \toprule
  &  Low intensity regime & High intensity regime \\[6pt]
\midrule
Value of friends of friends  & below threshold  $\tau_{\text{eq}}$ & above threshold $\tau_{\text{eq}}$ \\[6pt] 
Number of friends per agent & converging to a constant  & growing as $\sqrt{n}$ \\[6pt]
Connectedness & fragmented & fully connected   \\[6pt]
Diameter & $\infty$   & 3  \\[6pt]
Density depends on value of direct friends? & yes & no \\ \bottomrule  \end{tabular} \end{threeparttable}\end{center} \caption{A summary of the properties of the two equilibrium regimes; $n$ is the population size. The equilibrium falls into the low-intensity regime if the value of friends of friends is below the critical threshold $\tau_{\text{eq}}$, and into the high-intensity regime otherwise.} \label{tab:regimessummary} \end{table}

\begin{figure}[!t]
  \centering
  \subfloat[]{\includegraphics[width=3.5in]{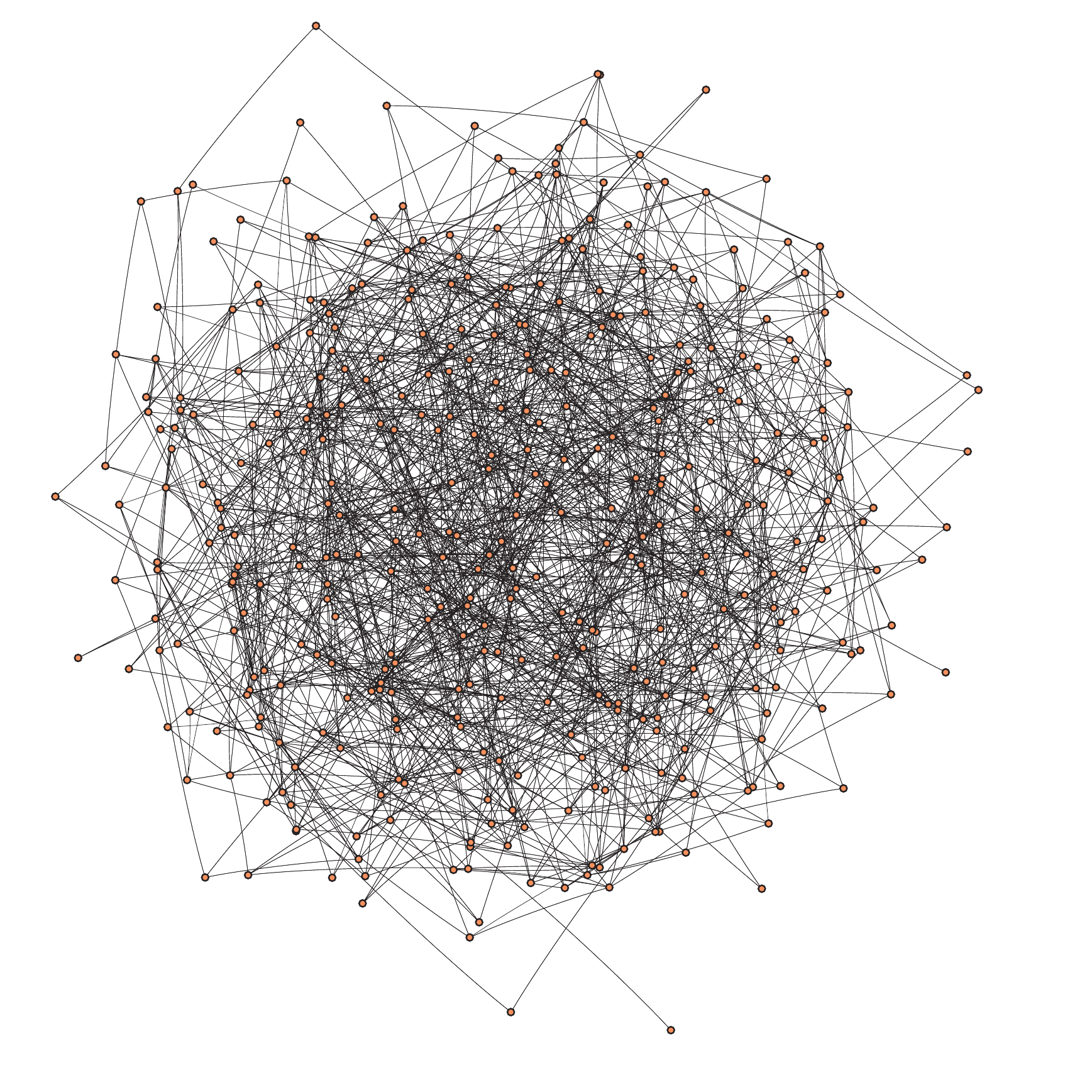}} \\
  \subfloat[]{\includegraphics[width=3.5in]{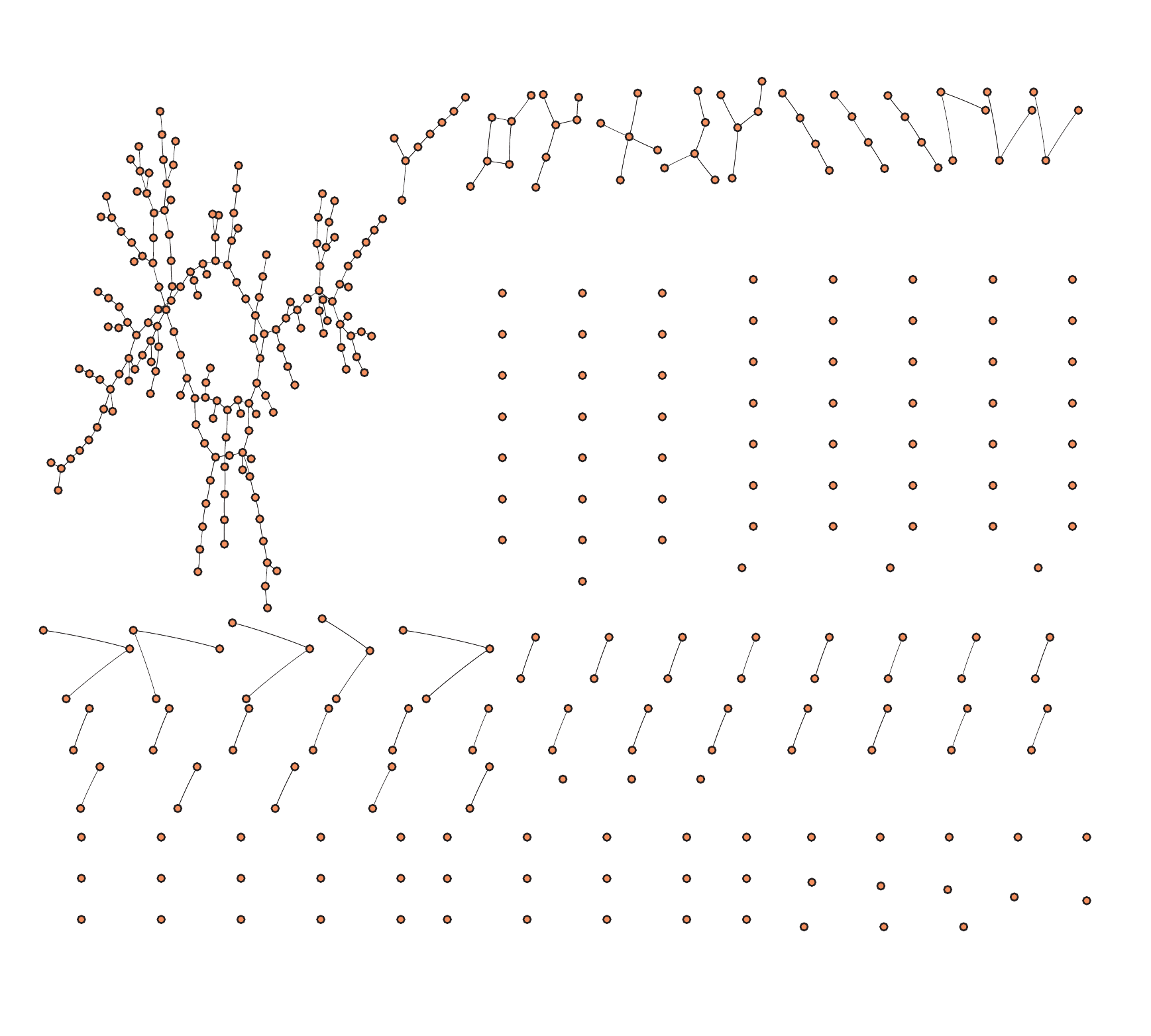}}\\
   \caption{Examples showing typical networks formed in equilibrium with $n=400$ agents in (a) the high-intensity regime and (b) the low-intensity regime. The high-intensity network has a single component and many links per node, whereas the low-intensity network is highly fragmented.}
  \label{fig:examples}
\end{figure}

The first main result is that equilibrium networks fall into two regimes, depending on the parameter values: a connected, high-intensity regime, and a fragmented, low-intensity one. These regimes are extremely different, and which one is relevant depends on the comparison of the value of friends of friends (an exogenous parameter called $v_2$) to a threshold called $\tau_{\text{eq}}$ that is computed based on costs. The properties of the regimes are summarized in Table \ref{tab:regimessummary}, and some illustrative examples are shown in Figure \ref{fig:examples}. When friends of friends are sufficiently valuable, with their value exceeding a cost-dependent threshold, agents in equilibrium devote a lot of time to socializing, and the expected number of friends each has scales as the square root of the population size. As already implied by the name, the networks in this regime are connected with very high probability as the population grows large --- indeed, there is a path of length at most three between any two agents. In contrast, when the value of friends of friends falls just slightly below the threshold, agents socialize significantly less, and the resulting networks consist of many disconnected pieces. The expected number of friends per agent tends to a constant as the network grows large. A striking fact is that the value of one's own friends affects neither the location of the transition $\tau_{\text{eq}}$ between these equilibrium regimes nor the properties of the connected, high-intensity regime. It only affects the equilibrium degree of each agent in the low-intensity regime.

The second main result focuses on efficiency. In the case where all agents have a known cost parameter, so that the social planner faces no informational problems, we characterize ``efficient mingling'': what uniform intensities of interaction a utilitarian social planner would select. We find that these, too, exhibit a phase transition when the value of friends of friends surpasses a key threshold $\tau_{\text{eff}}$, moving from low intensity to high intensity. However, this threshold governing the jump in \emph{efficient} levels of socializing happens at half the threshold governing the jump in the \emph{equilibrium} levels of socializing. That is, $\tau_{\text{eff}} = \frac{1}{2} \tau_{\text{eq}}$. Between these two thresholds, when $\tau_{\text{eff}} < v_2 < \tau_{\text{eq}}$, we find that efficient levels of socializing exceed equilibrium levels by a factor growing to infinity with the population size; the same is true of agents' welfare. Outside of this area between the thresholds, equilibrium networks are still inefficient, but only by a constant factor for arbitrary network size. That is, the area between the thresholds is one of extreme inefficiency, where a vanishingly small fraction of available benefits are being extracted. This is summarized in Table \ref{tab:efficiency}.

\begin{table}
\begin{center}
\begin{threeparttable}
\begin{tabular}{@{\extracolsep{.1in}}p{1.4in}p{1.5in}p{1.3in}p{1.3in}} \toprule
Value of friends of friends &  Equilibrium mingling intensity &  Efficient mingling intensity &  Welfare loss\tnote{a} \tabularnewline
\midrule
$v_2 \in (0, \tau_{\text{eff}})$ & low & low  & bounded \tabularnewline
$v_2 \in (\tau_{\text{eff}}, \tau_{\text{eq}})$ & low & high & unbounded \tabularnewline
$v_2 \in (\tau_{\text{eq}}, +\infty)$ & high & high & bounded \tabularnewline \bottomrule  \end{tabular}  \begin{tablenotes}
\footnotesize{\item[a] Relative to equilibrium welfare, for fixed costs and values, as the population size grows.}
\end{tablenotes}  \end{threeparttable}\end{center} \caption{Comparison between equilibrium and efficient networks, as well as how much of the available welfare is lost. When $v_2 \in (\tau_{\text{eff}}, \tau_{\text{eq}})$, the ratio of available gains to realized gains is unbounded, making this intermediate range extremely inefficient.} \label{tab:efficiency} \end{table}

These results can be interpreted in the context of social networking technology. When a new technology comes along that increases the value of friends of friends, for example by exposing lists of friends of friends and information about them, agents' equilibrium socializing decisions can shift drastically if the starting point was near the critical threshold. The technological shift takes the network from a fragmented and extremely inefficient regime to a connected and much more efficient one. This effect is only amplified by the fact that new technologies arguably also reduce the costs of socializing. Thus, our analysis provides a potential underlying mechanism for the large qualitative impact that social networking services have been said to have on social life.

The paper is organized as follows. In Section 1, we discuss how our approach relates to the economic literature on network formation. Next, in Section 2, we formally lay out the model. Then we examine equilibrium and efficiency --- first, in Section 3, when all agents have the same costs, and then, in Section 4, when their costs are private independent draws from a commonly known distribution. In Section 5, we discuss extensions of the results to highlight their robustness and limitations. There, we (i) endogenize the value of friends of friends through the mechanism of people introducing friends to each other; (ii) show that the specification of the cost function is not driving the qualitative results; (iii) explain why the realized random networks are stable when agents have to pay some costs to maintain links. Section 6 concludes.

\section{Related Literature}

The importance of the basic problem of how social networks form has been widely recognized in economics\footnote{Social networks affect economic outcomes in a multitude of ways. They influence decisions and outcomes relating to employment (Topa, \citeyear{topa}), investment (Duflo and Saez, \citeyear{duflo}), risk-sharing (Ambrus, Mobius, and Szeidl \citeyear{ams}), education (Calv\'{o}-Armengol, Patacchini, and Zenou, \citeyear{CPZ}), and crime (Glaeser, Sacerdote, and Scheinkman, \citeyear{glaeser}), to name just a few of their effects. See \cite{Granovetter_JEP} for a broad survey of the effects of social networks.}, and the study of rational network formation has a rich history. One strand of this literature, starting with \cite{myerson} and continuing with \cite{jackson-wolinsky-96}, \cite{BalaGoyal2000}, and \cite{szeidl}, among many others, has has studied the stability of certain networks to unilateral and bilateral deviations which translate deterministically into changes in the network. The literature is surveyed extensively by \cite{JacksonFormationSurvey} and \cite{JacksonBook}. This approach implicitly assumes that agents know the network insofar as that is important for their deviations, and delivers very specific and often stark predictions about network structure. While this has been an extremely important approach for understanding aspects of network formation, a different model is appropriate for the first-meeting setting that we focus on, as well as for generating the random graphs that are our equilibrium predictions. In our model, in contrast to these, any network has a positive probability of appearing in equilibrium, though some are much less likely than others; moreover, agents are fully aware of the randomness that generates this and take it into account when optimizing. This makes the present model a natural fit for structural estimation.

Another strand of the literature, which includes K\"{o}nig, Tessone, and Zenou \citeyearpar{formationZenou}, has modeled various dynamic processes of network evolution with rational choices; those models often include a stochastic aspect in how decisions translate into outcomes. The networks generated are thus random, but usually belong by construction to a specific sub-class of networks. This approach is similar to ours, though we focus on a simple static framework. The main distinction is that the random graphs that come out of our model are closer to the standard ones used in the statistical modeling of social networks.

The analysis closest to ours is a recent one by Cabrales, Calv\'{o}-Armengol, and Zenou \citeyearpar{conferencesZenou} --- henceforth CCZ --- which inspired this work. This paper also studies agents who choose how much to socialize, spreading their efforts uniformly within a group. The main difference is that in the CCZ model, these choices yield links of intermediate strength, so that if $i$ mingles with intensity $x_i$ and $j$ mingles with intensity $x_j$, then there is a link between them of strength $x_i x_j$ formed with certainty. In our model, there would be a link between them formed with \emph{probability} $p\left(x_i,x_j\right)$, where $p$ can be a general symmetric function satisfying very mild conditions; the presence of a link is discrete, so that it is either present or absent. This yields stochastic networks with richer structure, at the expense of more complex probabilistic calculations. The modeling differences arise from a difference in motivation: our focus is on agents who socialize for the purpose of forming future long-lived connections that may or may not be realized, while in the CCZ framework, the emphasis is on the spillovers that current socializing creates for current production. The models also yield different insights. Due to the smoothness of the interaction function, the CCZ approach permits a full-fledged equilibrium/welfare analysis of socializing and production.  In contrast, our approach takes a reduced-form view of network benefits but allows us to look at the surprising tipping points in mingling-based network formation and to model random networks with rational agents.

\section{The Environment} \label{sec:environment}

\paragraph{Basics} $n\geq 3$ agents want to form relationships (or ``links'') between them, and they start out unlinked. Agent $i\in\{1,..,n\}$ interacts with the other agents according to an overall \emph{intensity of interaction} $x_{i}\in [0,1]$, which is her choice variable. The \emph{quantity of interaction} between agents $i$ and $j$ is given by $p_{ij}=p_{ji}=p\left(x_{i},x_{j}\right)$, and a link will form in the future between agents $i$ and $j$ with probability equal to the quantity of interaction $p_{ij}$, independently across links. We assume that $p:[0,1]\times[0,1]\to [0,1]$ is a continuous symmetric function, which is strictly increasing on $(0,1] \times (0,1]$. Finally, we assume that $p(0,0)=0$ and $p(1,1)=1$, so that if both sides of a future relationship put no effort into it, then it has no chance of materializing, and if both are fully committed to it then it will succeed for certain.

\paragraph{Timing} In the first stage agents interact by setting their choice variables $x_i$; they pay the interaction costs up-front. At the second stage the social network is realized; we denote it by an $n$-by-$n$ symmetric matrix $\mathbf{G}$. The indicator variable of the presence of the link $\{i,j\}$ is written $G_{ij} = G_{ji} \in \{0,1\}$. Agents get benefits from having direct and indirect links in the \emph{realized} network.

\paragraph{Preferences} Agent $i$'s costs are convex in her total quantity of interaction, and explicitly take the form:
$$\frac{c_i}{2}\left(\sum_{j\neq i}p_{ij}\right)^2$$
where $c_i$ is an agent-specific coefficient capturing the cost of social interaction, which is the private information of agent $i$. It will be useful to refer to the inverse $c_i^{-1}$ as the \emph{sociability} of agent $i$, and denote it by $s$. Agent $i$ gains a value $v_1$ from any \emph{friend} (a $j$ such that $G_{ij} = 1$) and a value $v_2$ from each \emph{friend of a friend} -- an agent to whom she is not directly linked, but connected through at least one mutual friend.\footnote{An agent $j$ who is a friend of $i$ is \emph{not} called a friend of friend of $i$ even if there is a third agent $k$ who is linked to both $i$ and $j$. In terms of the utility function, this means that one does not get additional value from people one knows directly if one is also connected to them indirectly.} Formally, this is any $j$ so that $G_{ij} = 0$ but $G_{ik} G_{kj} = 1$ for some $j \neq k$.\footnote{One could also model agents as valuing contacts which are removed from them by two or more links in the realized network ($v_3,v_4,\ldots$). However, we do not view these as realistically having first-order effects on agents' considerations in network formation.} We assume that $v_1>v_2 \geq 0$. Thus, the expected utility of agent $i$ can be written as $$ u_i(\mathbf{x}) = \Ex\left[ v_1 \cdot \# \text{friends} + v_2 \cdot \# \text{friends of friends} \right] - \frac{c}{2}\left(\sum_{j\neq i}p_{ij}\right)^2.$$
 
\paragraph{Comments on the Modeling Assumptions} This is a model of social network formation among a large group of people, all meeting each other for the first time. People gain benefits from having friends and friends of friends in the future, yet have to spend costly time to form new relationships with others. A leading example is that of students coming into an MBA program. Some of the main benefits of pursuing an MBA are the social and professional connections that can be gained while getting the degree. However, students have only a limited amount of time during the program, and socializing displaces other valuable activities. Other suitable examples are provided by entering students in other academic programs, new recruits in the military, and businesspeople at a conference or trade show. As mentioned in the introduction, such situations display properties which are captured in the assumptions of our model, yet are very different from the ones underlying network-stability models, which take some existing network as a point from which to consider deviations. We now explain our assumptions in more detail in the context of the motivating examples.

First, the process of forming new relationships exhibits a substantial amount of fundamental uncertainty, in contrast to the maintenance of existing relationships. When two people have known each other for a long time and are willing to make the investment required to continue the relationship, it is most likely that the link between them will prevail. In contrast, we view the process of forming new relationships as fundamentally uncertain. New acquaintances may not maintain a relationship for many reasons: they might not share enough common interests; they may move to different locales before the relationship is established; they may realize they do not like each other; or they might simply lose touch because of exogenous distractions. Thus, the model features probabilistic network formation. When investing effort in socializing, agents can affect the probability of a relationship forming but cannot guarantee it unless they both invest the maximum possible amount. Otherwise, our model allows for a great deal of generality in how decisions translate into relationship probabilities.

Second, when investing in the formation of a link with other agents, the major cost involved is the time it takes to get to know him or her and establish a basis for a potential relationship. A substantial portion of this time is spent in the beginning stages of the relationship, while benefits from the relationship may be reaped over many years in the future. Thus, costs in our model are paid up-front, before the structure of the network is determined, while utility from the social network is accrued at the second stage, after links are realized. Accordingly, we initially assume away costs of maintaining links after they have formed. In Section \ref{sec:stability} we introduce maintenance costs into our model and show that they do not affect our main results. 

Third, like \citeyearpar{conferencesZenou}, we assume that agents cannot target specific individuals when socializing; instead, they interact generally within the group. In the MBA example, this would be equivalent to students deciding on how many parties to go to or MBA clubs to join. In Section \ref{sec:mingling} we relax this assumption and show that for reasonably large populations and in the presence of some decreasing marginal returns to socializing, this assumption is in fact a result obtained in equilibrium.

Fourth, we make the standard assumption that friends of friends are valuable. This is consistent with previous models. However, we mention it because it is a key driving force in the analysis. At first we will work under that assumption that this benefit from having friends of friends is an exogenous parameter, and will later endogenize it.

Fifth, we assume that agents time costs are a function of the sum of their probabilities of forming relationships. The two essential substantive assumptions behind this are the ``independence'' of the interactions across pairs of agents, and the ex-ante symmetry of all pairs in terms of how costly interaction is.

\section{Homogeneous Agents}

We now turn to analyzing the model, first focusing on equilibrium behavior of the strategic agents, and then on efficiency. In this section we assume that all agents are homogeneous, in the sense that they all have the same costs of social interaction time $c$, and in the next section this assumption will be relaxed. All proofs are provided in the appendix.

\subsection{Equilibrium} The solution concept we use is a \emph{symmetric equilibrium} in which all agents choose the same intensity $x$. Since we are modeling a process of network formation, where agents start out without any prior information about each other, or regarding some target or benchmark social network, any equilibrium concept which is asymmetric implicitly assumes some mechanism for coordination. In the absence of such a mechanism, a symmetric equilibrium is a particularly compelling concept in our setting.

In this game there always exists a trivial symmetric equilibrium -- the one in which all agents socialize with intensity 0 with everyone else. Our first result establishes the existence and uniqueness of a nontrivial symmetric equilibrium, in which every agent plays her unique best response.

\begin{theorem} \label{thm:existence}
A symmetric equilibrium with positive linking probabilities always exists, and is unique. This is a strict equilibrium, in the sense that each agent has a unique best response.
\end{theorem}

We will denote the probability of a relationship forming between any two agents in a symmetric equilibrium by $p^*$. While we can only characterize $p^*$ implicitly, this allows us to provide comparative statics with respect to the costs and benefits from direct and indirect friends. 

\begin{proposition} \label{prop:compstat}
Let $p^*$ be the probability of forming relationships in the symmetric equilibrium. If $p^*<1$, it holds that:
	\begin{enumerate}
	\item 	$\frac{\partial p^*}{\partial c}<0$
	\item 	$\frac{\partial p^*}{\partial v_1}>0$
	\item 	For a fixed $n$, there exists a threshold probability $\hat{p}(n)>0$ such that if $p^*<\hat{p}$ then $\frac{\partial p^*}{\partial v_2}<0$, and if $p^*>\hat{p}$ then $\frac{\partial p^*}{\partial v_2}>0$.
	\end{enumerate}
\end{proposition}

The first two comparative statics we present are relatively straightforward. With rising costs agents find it more costly to interact and thus reduce their equilibrium intensity of interaction. Similarly, increasing the benefits from direct friends directly increases the personal benefits from interaction, and hence the equilibrium probability of forming relationships rises.

The comparative statics with respect to the value of friends of friends are more subtle. While an increase in the value of friends of friends does render interaction more valuable, it also promotes a free-riding incentive, since an agent does not pay directly for her friends of friends. Holding the other parameters fixed, there is a critical level of $p^*$ such that for equilibrium probabilities above it the free-riding effects is stronger, and for probabilities beneath it the threshold the direct effect triumphs. The intuition is the following: with very high equilibrium linking probabilities agents form relatively many links, and thus expect to have relatively many friends of friends, with frequent overlap --- an agent will tend to know friends of friends through multiple direct friends. When $v_2$ rises, the free-rider incentive to reduce social interaction intensities and cutting costs are thus first order and trump the increased value in more interaction. When $p^*$ is very low, the network is sparse and agents expect few friends of friends, and thus the dominating effect is the direct one.

\subsection{Asymptotic Results} Since we are modeling large social networks of hundreds or thousands of agents, it is instructive to analyze the asymptotic behavior of the equilibrium when $n$ grows large. Our main result shows that this behavior can take on two very different forms. To state it formally we define $F_i$ to be the expected number of friends for agent $i$ in the unique symmetric equilibrium, where $F_i=(n-1)p^*$. 

\begin{theorem} \label{thm:maintheorem}
The network formation symmetric equilibrium is asymptotically governed by two possible regimes:
	\begin{enumerate}
		\item 	If $v_2<\tau_{\text{eq}}=c$, then the equilibrium linking probability decays at a rate of $1/n$ and the expected number of friends each agent has in equilibrium converges to a finite number: $$\lim_{n\to\infty}F_i=\frac{v_1}{c-v_2}$$
		\item 	If $v_2>\tau_{\text{eq}}$, then the equilibrium linking probability decays at a rate of $n^{-\frac{1}{2}}$ and the expected number of friends each agent has in equilibrium grows to infinity at a rate of $n^{\frac{1}{2}}$:
$$\lim_{n\to\infty}n^{-\frac{1}{2}}F_i=\log^{\frac{1}{2}}\left(\frac{v_2}{c}\right)$$
	\end{enumerate}
\end{theorem}

Theorem \ref{thm:maintheorem} shows that equilibrium behavior, asymptotically, can fall into two starkly different regimes. The governing regime depends on the comparison between $v_2$, the value of friends of friends, and the cost coefficient $c$, such that when $v_2$ crosses over the threshold $\tau_{\text{eq}}=c$, a phase transition occurs. In the \emph{high-intensity regime}, which governs when $v_2$ is high relative to the cost coefficient $c$, agents invest long periods of time interacting with one another, and the expected equilibrium number of friends grows with the size of the population, $n$, at a scale equal to the square root of $n$. Intuitively, this expected number of direct friends each agent makes in equilibrium, or expected degree, is increasing in the value of indirect friends, and decreases in costs. Surprisingly, the asymptotic degree does not depend at all on the value of direct friends. This implies that in the high-intensity regime the dominant factor compelling agents to interact is the benefit of indirect friends. Since the number of friends of friends is, approximately, the square of the number of friends, in the limit this benefit completely trumps out the direct benefit of friends.

In contrast, in the \emph{low-intensity-regime}, when $v_2$ is low relative to $c$, agents interact at an intensity which is an order-of-magnitude less than in the high-intensity-regime. The expected number of friends each agent makes in equilibrium now tends to a constant as the population grows large. This constant is an increasing function of the value of both friends and friends of friends, but while it is a linear function of $v_1$, it is a non-linear function of $v_2$, which explodes when $v_2$ draws close to the threshold  $\tau_{\text{eq}}$ from below. Intuitively, the asymptotic expected number of friends is a decreasing function of the cost coefficient. In contrast with the high-intensity regime, the value of direct friends has a first order effect on the equilibrium intensities since in the low-intensity regime the number of friends of friends does not completely overpower the number of direct friends. 

The qualitative difference between the local network behavior in the two regimes results in dramatic differences in overall features of the entire network. To describe these differences we shall define the following terms: we say that a network $\mathbf{G}$ is \emph{connected}, if for any two agents $i,j$ there exists a sequence of agents $i_1,\cdots,i_n$ linking them, such that $G_{i,i_1}=1$, $G_{i_n,j}=1$, and for every $1\leq k \leq n-1$, $G_{i_{k},i_{k+1}}=1$; we say that agents $i,j$ are at \emph{distance} $k$ in a network $\mathbf{G}$ if the shortest path connecting them in $\mathbf{G}$ is of length $k$; finally, given a network $\mathbf{G}$, we define the \emph{diameter} of $\mathbf{G}$ to be the maximum distance between any two agents in $\mathbf{G}$.

Using classical results from the theory of random networks, we characterize the network-level differences between the two regimes in the following result. We say a statement holds ``asymptotically almost surely'' (a.a.s.) if it holds with a probability that tends to $1$ as $n$ grows.

\begin{proposition}\label{prop:properties}
In the high-intensity regime the realized social network is connected asymptotically almost surely, and the diameter of the network converges in probability to 3.

In the low-intensity regime the realized network is a.a.s. not connected. If $\frac{v_1}{c-v_2}>1$ the network will a.a.s. contain a single connected component which includes a positive fraction of the agents, and if $\frac{v_1}{c-v_2}<1$ the network will a.a.s. not contain any component larger than $O(\log(n))$ agents.
\end{proposition}

This result shows that the difference between high-intensity and low-intensity regimes yields sharp empirical predictions at the macro-network level. High-intensity regime networks are connected with a very high probability, so that every any agent is linked, directly or indirectly, to any other agent. Moreover, with a probability that tends to 1, any two agents are at most three friends away from each other, and there exists a pair of agents that are exactly three friends separated from each other.

Low-intensity networks on the other hand tend to be disconnected. Depending on the size of the average degree $\frac{v_1}{c-v_2}$ relative to 1, these networks may display a giant component, which connects a positive fraction of the agents to each other. If such a component does not exist, the social network tends to display connected components which are tiny when compared to the overall size of the network.

Perhaps the most striking feature of the two regime structure is that the transition between the two regimes does not depend on the value of direct friends $v_1$, but only on the cost coefficient $c$ and the value of friends of friends, $v_2$. This implies the moving from one regime to the other depends on the burden of personal costs on the one hand, but on the other only on those benefits that come from the network structure itself which the agent does not control --- the indirect benefit of friends of friends.

Moreover, this characterization of the threshold implies that for some networks, small changes in either the costs of social interaction or the in the value of friends of friends can cause dramatic differences in the resulting social networks. In light of Proposition \ref{prop:properties}, these differences will manifest both at the individual level, as the average agent will have many more friends, yet also in the macro-network level. As we will show below, this shift is also associated with a sharp rise in efficiency.

\subsection{Social Networking Technology} This result can shed some light on the recent developments in social networking technologies, and specifically the massive rise and arguably substantial impact of online social networks such as Facebook, MySpace, LinkedIn and Twitter. Hundreds of millions of people now use these networks regularly, spending, on average, long periods of time using them daily (Boyd and Ellison, \citeyearpar{boyd:sns}; ``Facebook: Statistics'' \citep{facebook}). While these networks offer their user different and perhaps easier forms of connecting with friends, the direct benefits of using them (to browse photographs, exchange messages, etc.) are arguably similar to other technologies whose impact has been less dramatic. It is clear though, that these networks specifically and intentionally increase users' benefits from indirect friends. All of the above networks expose a user to the identity of friends of friends, usually providing some information about them, such as occupations, photos, hobbies and interests. Moreover, some of these tools, like LinkedIn, explicitly emphasize the value they add to friends of friends by showing users how they can connect to certain individuals or organizations through their personal and professional social network. In light of this, our model suggests one answer to the question of the real value underlying the success of online social networks --- by increasing the benefits of indirect friends in real social networks, while also arguably reducing the costs of interaction, they may be pushing the formation of social networks beyond the critical threshold and into the high-intensity regime.

\subsection{Efficient Symmetric Socializing} Returning to our model, we can ask within the same framework how far the symmetric equilibrium network is from the efficiency --- a network with a social planner choosing for each agent an optimal level of interaction intensity. Intuitively, our model is characterized by only one externality, as strategic agents do not internalize the benefits of acquiring more friends would have on other acquired friends through benefits from friends of friends. Thus, we would expect the equilibrium intensity levels to be inefficiently too low. Our next result confirms this, but also sheds light on the size of this inefficiency.

\begin{figure}[!t] \begin{center}
\includegraphics[width=6.4in]{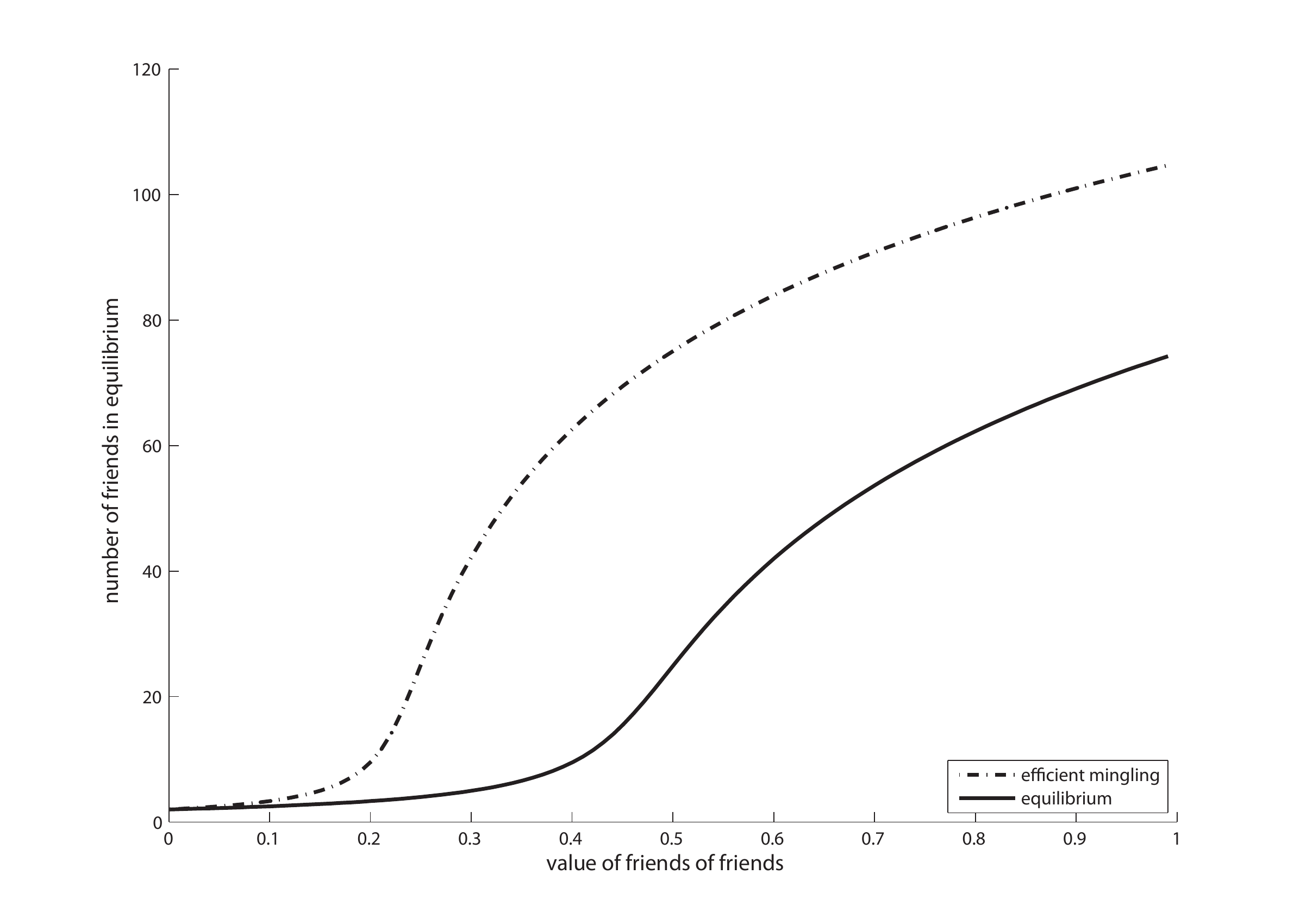}
\caption{The equilibrium and efficient amounts of socializing (reflected in numbers of friends) as $v_2$, the value of friends of friends, is varied for $n=8000$ agents. The threshold at which the efficient levels of socializing transition into the high regime $(\tau_{\text{eff}} = .25$) is half the threshold at which the equilibrium levels do $(\tau_{\text{eq}} = 0.5)$. Between these thresholds, the resulting network is very inefficient. (The cost parameter is $c=0.5$, and the value of friends is $v_1 = 1$.)} 
\label{fig:efficiency}
\end{center}
\end{figure}

\begin{theorem} \label{thm:efficiency}
The socially optimal level of linking probabilities is strictly higher than the equilibrium one. Efficient interaction is governed by two regimes, separated by a threshold $\tau_{\text{eff}}=\tau_{\text{eq}}/2$:
\begin{enumerate}
	\item 	If $v_2<\tau_{\text{eff}}$ then the efficient linking probability, $\hat{p}$, decays at a rate of  $1/n$ and the efficient expected number of friends each agent has converges to a finite number: $$lim_{n\to\infty}F_i=\frac{v_1}{c-2v_2}$$
	\item 	If $v_2>\tau_{\text{eff}}$, then the efficient linking probability decays at a rate of $n^{-\frac{1}{2}}$ and the efficient expected number of friends each agent has grows to infinity at a rate of $n^{\frac{1}{2}}$:
$$\lim_{n\to\infty}n^{-\frac{1}{2}}F_i=\log^{\frac{1}{2}}\left(\frac{2v_2}{c}\right)$$
	Under this regime, agents' welfare tends to infinity.
\end{enumerate}
\end{theorem}

We thus have that efficient intensities behave similarly to equilibrium levels and fall into one of two regimes which according to the ratio between $v_2$ and $c$, once again independently of the value $v_1$. However, as expected, equilibrium linking probabilities are lower than efficient. Perhaps surprisingly, these inefficiencies do no disappear in the limit. For very low $v_2$, such that $v_2<\tau_{\text{eff}}$, and for high $v_2$, such that $v_2>\tau_{\text{eq}}$, the asymptotic equilibrium probability is a fixed fraction of the asymptotic efficient probability. The same holds for agents' welfare. However, for intermediate values of $v_2$, such that $\tau_{\text{eff}}<v_2<\tau_{\text{eq}}$, the inefficiency becomes extreme, and efficient probabilities and welfare are infinitely larger than equilibrium ones --- in equilibrium each agent has only a finite number of friends, while the efficient number of friends goes to infinity, along with welfare. In the social networking technology context, this implies that the networks for which a small changes in $v_2$ can promote a phase transition in interaction, are exactly those networks suffering the most in terms of efficiency. Thus, the importance of social networking technologies is further emphasized. This phenomenon in a particular example is illustrated in Figure \ref{fig:efficiency}.

\section{Heterogeneous Agents}

We now extend the model to allow for unobserved heterogeneity between agents. Specifically, we assume that agent $i$'s cost coefficient $c$ is drawn independently according to a probability vector $\textbf{p}$ over a finite set of possible costs $\mathcal{C}=\left\{c_1,...,c_m\right\}$, and that the realization of this draw is agent $i$'s private information. We assume that the pair $(\textbf{p},\mathcal{C})$ is commonly known. The different cost coefficients may represent two differences between the agents: First, agents may differ in how easy it is for them to interact and spend time with others; Second, agents may have different marginal costs of time due to different alternative uses they have for that time. We will denote by $\Ex\left[C\right]$ the expectation under $\textbf{p}$ of the cost coefficient, and similarly by $\Ex\left[S\right]=\Ex\left[1/C\right]$, the expectation of sociability. 

For simplicity, we assume in this section a specific functional form for the function linking interaction intensities and link realization probabilities: $p(x,y)=xy$. This functional form satisfies all our original assumptions on the function $p$. Although we make this specific choice for $p$, the results of this section generalize to a wide class of functions.

In this context, we continue to use a symmetric equilibrium as the solution concept, where any two agents with the same realized costs use the same intensities. We first establish that a symmetric equilibrium exists in this extended context, and that it is still a strict equilibrium.

\begin{theorem}\label{thm:hetroexistence}
There exists a symmetric equilibrium with strictly positive interaction intensities. In this equilibrium agents use their unique best responses.
\end{theorem}


We next turn to characterize the asymptotic behavior of equilibrium relationship realization probabilities. Perhaps surprisingly, the increased complexity of the environment does not change the qualitative nature of the asymptotic convergence when compared with the homogeneous case. However, the threshold separating the two regimes takes on a more complicated form, which depends on the distribution of costs.

\begin{theorem} \label{thm:hetro}
The symmetric equilibrium with heterogeneous, private costs, is asymptotically unique, and governed by two possible regimes which are separated by a threshold $\tau_{\text{eq}}=\Ex\left[S \right]/\Ex\left[S^2\right]$:
	\begin{enumerate}
		\item 	If $v_2<\tau_{\text{eq}}$, then the equilibrium linking probabilities decay at a rate of $1/n$ and the expected number of friends each agent has in equilibrium converges to a finite number, so that for agent $i$ with private cost $c_i$: $$\lim_{n\to\infty}F_i=\frac{s_iv_1}{\Ex\left[S\right]-v_2\Ex\left[S^2\right]}$$
		\item 	If $v_2>\tau_{\text{eq}}$, then the equilibrium linking probabilities decay at a rate of $n^{-\frac{1}{2}}$ and the expected number of friends each agent has in equilibrium grows to infinity at a rate of $n^{\frac{1}{2}}$, and is independent of $v_1$.
	\end{enumerate}
\end{theorem}

In analogy with Theorem \ref{thm:maintheorem}, we have that with heterogeneous agents holding private information, the symmetric equilibrium can behave according to two regimes. In the high-intensity regime, which governs when the $v_2$ is higher than the threshold $\tau_{\text{eq}}$, the expected equilibrium number of friends agents make, regardless of their private costs, converges to infinity at a rate of $\sqrt{n}$. The exact rate of convergence is characterized implicitly in the proof of Theorem \ref{thm:hetro}, and is independent of $v_1$. Following the same arguments as in the homogeneous case, the resulting network in the will be connected, and the diameter of the network will be three, asymptotically almost surely. Perhaps surprisingly, these results hold regardless of the exact nature of the distribution of costs in the population.

In the low-intensity regime, which governs when $v_2$ is higher than the threshold $\tau_{\text{eq}}$, the expected equilibrium number of friends converges when $n$ grows large to a fixed number, which is proportional to the agents' sociabilities. The asymptotic expected number of friends is increasing linearly with $v_1$ in this regime, and increasing non-linearly in $v_2$, exploding when $v_2$ is close to the threshold $\tau_{\text{eq}}$ from below. As in the homogeneous case, the resulting network will be disconnected asymptotically almost surely, and the existence of the giant component will depend on the comparison between the expected equilibrium number of friends and 1.

The threshold level itself, $\tau_{\text{eq}}$, is the ratio between the first two moments of the distribution of costs. Rewriting this threshold as: $$\tau_{\text{eq}}=\frac{\Ex\left[S\right]}{\Ex\left[S\right]^2+\Var\left[S\right]}$$ yields the following comparative statics.

\begin{corollary}
$\tau_{\text{eq}}$ decreases with mean-preserving spreads of the distribution of sociabilities. A variance-preserving increase in the mean of the distribution of sociabilities will increase $\tau_{\text{eq}}$ when the mean is low compared with the standard deviation, and will decrease it otherwise. 
\end{corollary}

The first comparative static suggests that the presence of some agents with higher sociabilities, or lower cost coefficients, even at the expense of agents with lower sociabilities, can be crucial for a forming group to reach the beneficial high-intensity regime. This implies that highly sociable individuals, which will in equilibrium obtain many friends, and this will provide their friends with many indirect links, may be the key for a highly connected group.

The second comparative static shows, perhaps surprisingly, that increasing the sociability of all agents in society by the same amount can be detrimental to transition into the high-intensity regime, and this is when the mean sociability is high relative to its standard deviation. The intuition for this counter-intuitive result is free-riding --- when all agents' sociability increases, agents with low sociabilities receive a higher increase in percentage terms. Agents with higher sociabilities, who are not that far apart because of the relatively low variance, will thus have a greater incentive to free-ride, and this might trump the direct effect of lowered costs.

\section{Extensions} \label{sec:extensions}

\subsection{Allowing Discrimination: Mingling as an Equilibrium} \label{sec:mingling}

In the description of our game, we assumed that agents choose one intensity for socializing within the group in general, without the possibility of discriminating. While this can be motivated as a reasonable restriction based on the difficulty of coordinating and focusing on specific others at the early stages of interactions, as in Cabrales, Calv\'{o}-Armengol, and Zenou \citeyearpar{conferencesZenou}, we do not have to view this as a restriction. Indeed, we can enrich the model to one in which each agent $i$ chooses an intensity $x_{ij}$ to direct at each other agent $j$, and the probability that they link is the quantity of interaction $p_{ij} = p\left(x_{ij}, x_{ji}\right)$. The rest of the model is unchanged.

When the baseline model is enriched in this way, mingling --- devoting equal intensity to every other agent --- is not an equilibrium. However, if we add a small and realistic perturbation to the theory --- namely, by supposing that the probability of forming a friendship is slightly \emph{concave} in the quantity of interaction --- mingling becomes a strict equilibrium. The predictions of that nearby model --- in terms of the mingling intensities and the networks that are formed --- are close to the baseline model. In this section, we detail these points.

First, it is useful to understand precisely why mingling is not an equilibrium. If everyone else were mingling, agents would prefer to deviate and focus their efforts only on some subset of others. The feature driving this phenomenon is the fear of overlap between the neighborhoods of friends. For an illustration of this, suppose that an agent $i$ has only two potential friends $j$ and $k$, and each of them will have $3$ friends, in expectation, in addition to $i$. In expectation, there will be one agent other than $i$ who is a friend of both $j$ and $k$. Suppose also that $j$ and $k$ both direct intensities $x_{ji} = x_{ki} = y$ at agent $i$. If agent $i$ fixes the sum of intensities $x_{ij} + x_{ik}$ and decides how to apportion his socializing, it is straightforward to verify that he prefers to focus on one agent, either $j$ or $k$. This is because of the convexity of benefits introduced by the overlap: linking to both is less than twice as good as linking to a single one.

This overlap problem is a minor artifact rather than a major consideration because the amount of overlap in equilibrium will be relatively tiny. To make this precise, suppose we introduce a small amount of concavity in the function that maps total time spent together to the probability of a link forming. Equivalently, via a reparameterization, let us assume that the costs of interaction time are slightly convex, so that the utility function takes the form $$ u_i(\mathbf{x}) = \Ex\left[ v_1 \cdot \# \text{friends} + v_2 \cdot \# \text{friends of friends} \right] - \frac{c}{2}\left(\sum_{j\neq i}p_{ij}^\beta\right)^2$$ for some $\beta > 1$ --- where we have, of course, chosen a simple form of convexity.

In this framework, it can be shown that, for any $n$, there is a range $[\underline{\beta},\overline{\beta}]$ such that, when $\beta$ is in this range, mingling is a strict equilibrium, and the mingling intensity is within a specified distance (say, $10\%$) of that of the baseline model. While this range will depend on $n$, our preliminary calculations show that it is reasonably large for the values of $n$ where the theory would be relevant --- populations of sizes between several dozen and several thousand. The precise calculations will be included in a later draft.

The intuition, however, is simple. Mingling imposes only a mild constraint on the agents because overlap in friends' neighborhoods is tiny. Adding a slight amount of decreasing marginal returns is enough to make mingling strictly the best response to mingling

\subsection{Perturbations of the Utility Function}

While the linear-quadratic specification is standard and obviously advantageous from the standpoint of tractability, one might naturally wonder about whether the two regimes and other stark features of the model are driven by the particular parameters chosen for the analysis. The answer is that while the specification and the asymptotics we have focused on are useful tools for analysis and exposition, the actual predictions about equilibrium behavior for finite populations are robust to perturbations in agents' utility function.

\begin{figure}[!t] \begin{center}
\includegraphics[width=6.4in]{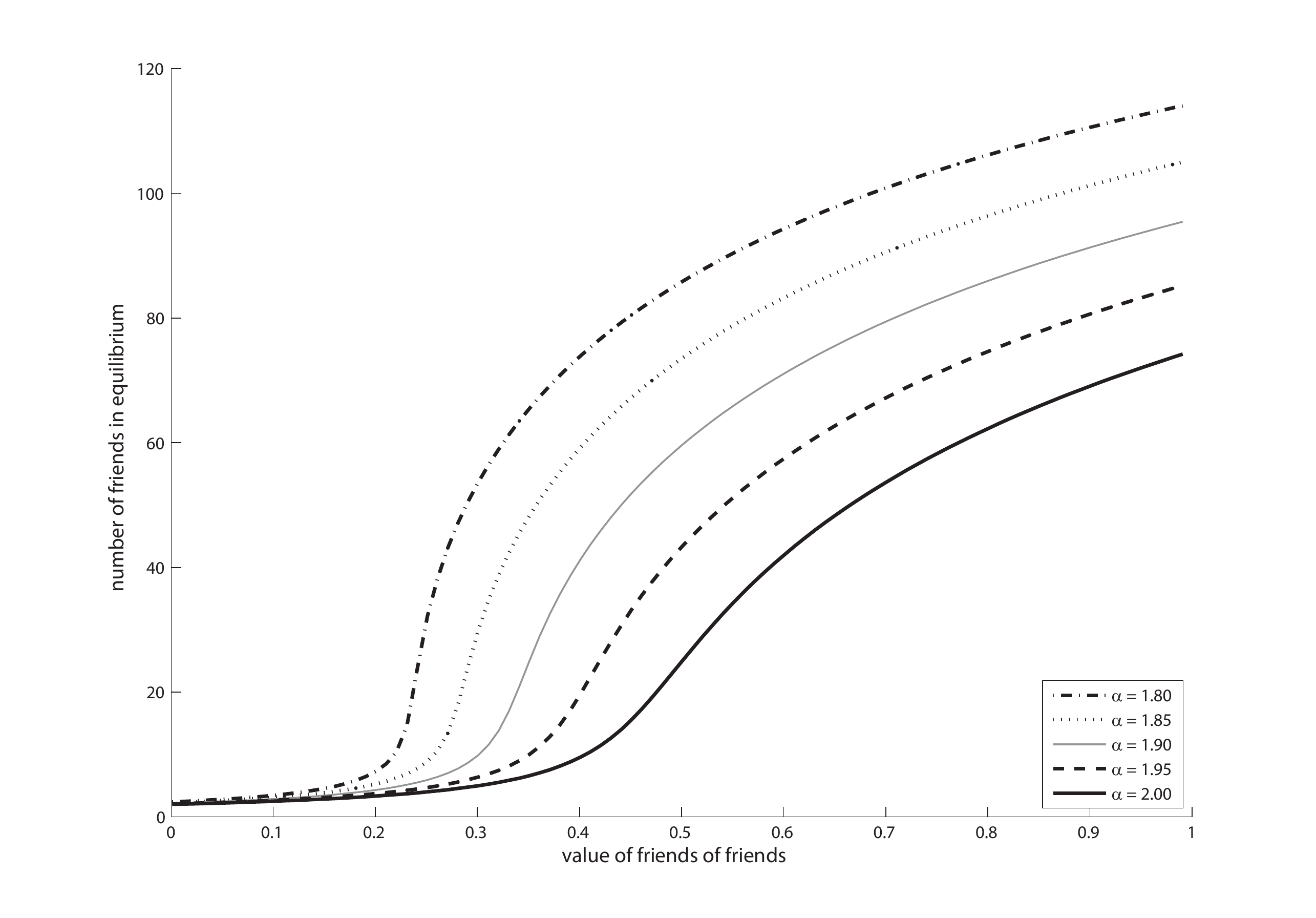}
\caption{Equilibrium numbers of friends when the model is perturbed and the costs of time scale as time to the power $\alpha$. In the baseline model, $\alpha = 2$. The numerical calculations in this network of $n=8000$ agents shows that the same basic pattern identified in the asymptotic analysis continues to hold. As costs grow less convex, the threshold at which the equilibrium shifts to the high-intensity regime occurs at a lower value of $v_2$. (The cost parameter is $c = 0.5$ and the value of friends is $v_1 = 1$.)} 
\label{fig:varyalpha}
\end{center}
\end{figure}

To explore this issue, we consider generalizing the utility function to: $$ u_i(\mathbf{x}) = \Ex\left[ v_1 \cdot \# \text{friends} + v_2 \cdot \# \text{friends of friends} \right] - \frac{c}{\alpha}\left(\sum_{j\neq i}p_{ij}\right)^\alpha,$$ where $\alpha > 1$ is a parameter. This captures the notion that costs of socializing might not scale exactly quadratically in total quantity of itneraction.\footnote{This can also be seen, via a reparameterization, as varying the concavity of benefits as opposed to the convexity of costs.}

If $\alpha > 2$, then arguments similar to those in the appendix show that the low-intensity regime is the only one that survives asymptotically; our numerical calculations for finite populations indicate that, indeed, intensities in equilibrium are low in finite populations. When $\alpha < 2$, \emph{asymptotically} the high-intensity regime is seen over the whole range of cost and benefit parameters. However, in finite populations, we still see a reflection of the same basic pattern identified by our asymptotic results with $\alpha = 2$. When the value of friends of friends is very low, equilibrium intensities are low, but they rapidly shift to a high level when $v_2$ surpasses a critical level. The main difference is that as $\alpha$ is reduced, that threshold occurs earlier. This is illustrated in Figure \ref{fig:varyalpha}.

The basic message is that working with $\alpha = 2$ and the asymptotic regime is a convenient expository device. It helps us identify the two regimes in a simple way and characterize their properties. But the numerics show that in reasonably large finite populations, less convex cost functions give qualitatively similar equilibria. Characterizing the cutoffs between regimes exactly as functions of $n$ for values of $\alpha$ other than $2$ would be an interesting extension of the analysis.

\subsection{Maintenance Costs and the Stability of the Random Networks} \label{sec:stability}

For simplicity of exposition, we have focused on the up-front costs of link formation, and have ignored the costs of maintaining links. Such costs introduce stability considerations; it may be that an agent does not wish to maintain a link once it has formed, because the link requires more effort to maintain than the marginal benefit it returns. We say that a network is \emph{unilaterally stable} if no agent wants to sever a link for this reason.

It turns out that maintenance costs are simple to incorporate into the model, and there is a reasonable range of parameter values for which the networks formed in the equilibrium of the formation game are stable asymptotically almost surely. To be precise, suppose that, to maintain a link and receive the benefits, an agent must pay a cost of $\tilde{c}$ in addition to the up-front costs. Here costs and benefits are interpreted as flows to be received in the future. We say a network is not unilaterally stable if there is some agent $i$ and some link incident to $i$ so that deleting the link results in a decrease in the network-based benefits to $i$ (from friends and friends of friends) of less than $\tilde{c}$.

A link to agent $j$ confers a marginal benefit of at least $v_1 - v_2$ upon agent $i$. This is because there may be some $k$ who is linked to both $i$ and $j$, so that if $i$ severs the link to $j$, she will still receive a benefit $v_2$ from being indirectly connected to $j$. Thus, if $\tilde{c} < v_1 - v_2$, any network that forms will be unilaterally stable. The following proposition says that, in the low-intensity regime, this condition is not only sufficient, but also necessary, for the formed network to be unilaterally stable asymptotically almost surely. That is, if $\tilde{c} > v_1 - v_2$, then  some agent will wish to sever a link with a probability bounded away from $0$ even as $n$ grows very large. 

\begin{proposition} In the low-intensity regime, the network created by the formation game is a.a.s. unilaterally stable if and only if $\tilde{c} \leq v_1 - v_2$. \label{prop:unistable} \end{proposition}

The proposition shows that, in general, when $\tilde{c} > v_1 - v_2$, the endogenous severing of links after the network is formed is a live possibility in at least one regime, which substantially alters the agents' expected utilities at the formation stage because they have to account for the possibility that some of their links will vanish. The analysis of the ways in which agents' formation-stage calculations change as a result of this possibility would be an interesting direction for further study.\footnote{In the high-intensity regime, we conjecture that the network is a.a.s. unilaterally stable for any finite $\tilde{c}$. This is because each friend of $i$ provides indirect connections to a large number of agents whom $i$ would not have access to through other friends. We also conjecture that allowing $v_1$ to be large and negative to account for agents foreseeing the maintenance cost would not affect the analysis of that regime.}

 On the other hand, when $\tilde{c} < v_1 - v_2$, stability considerations don't introduce any strategic complexity, because no links are severed after they are formed. The only issue is that agents' expected benefits from a direct friendship change from $v_1$ to $\tilde{v}_1 = v_1 - \tilde{c}$, because they will have to pay this maintenance cost.\footnote{We make the inequality $\tilde{c} < v_1 - v_2$ strict in order to ensure that $\tilde{v}_1 > 0$, which is assumption that we have maintained throughout the analysis of the formation stage.}  Subject to this replacement, the theory goes through unchanged.
 
 In short, while stability considerations can introduce substantial complexities into the analysis, there is a range of maintenance costs  in which the networks formed in equilibrium are a.a.s. robust to the possibility of unilateral deviation and agents' formation-stage accounting for the maintenance costs of links changes is straightforward to incorporate into the basic model.
 
\section{Concluding Remarks} \label{sec:conclusion}

This model of network formation with rational agents and uncertainty in the realization of links has two main appealing features. First, the networks it predicts have the complex and irregular structure seen in real networks \citep{NewmanSurvey}; moreover, they correspond to the random network models recently developed in the probability literature \citep{chunglu, ChungLuVu}. At the same time, the model does not rely on mechanistic
foundations for link formation; the probabilities of links are endogenous choice variables that are selected when agents optimize, trading off the costs of socializing against the expected benefits. From a technical perspective, the fact that there is uncertainty over the precise realizations of the links, along with convex costs of socializing, suffices to pin down equilibrium choices, in contrast to models of network formation where there is a multiplicity of equilibria.

The main results of the paper serve as an illustration of the ways in which the simple framework can generate nontrivial predictions about how the economic fundamentals affect equilibrium and efficiency. In the particular application considered here, we showed that small changes in the value of friends-of-friends can change the orders of growth of social activity and the fundamental shapes of equilibrium networks. The framework is capable of accommodating other specifications of costs and benefits -- for instance, ones that depend on how many mutual friends are between two agents, or on properties such as transitivity. 

While we made some first steps toward connecting rational random network formation and the stability of networks in Section \ref{sec:stability}, much remains to be done in exploring which formation processes result in persistent networks. This issue becomes more prominent when larger classes of deviations are admitted or when the network-based benefits take a more complex form, and would be an interesting direction for further research.

\newpage

\section*{Appendix}
\small
\paragraph{Expected Number of Friends of Friends} A calculation that will come in handy for most of the proofs is explicitly determining the expected number of friends of friends for agent $i$, given the intensity levels $\left\{x_j\right\}_{1\leq j\leq n}$. We claim it is the following expression:
$$\sum_{k\neq i}\left(1-p(x_{i},x_{k})\right)\left(1-\prod_{l\neq i,k}(1-p(x_{i},x_{l})p(x_{l},x_{k}))\right)$$
First, the index $k$ sums over all possible friends of friends $k\neq i$. For $k$ to be a friend of a friend, she needs to not be a direct friend, but to be a friend of some direct friend of $i$. The first term in the summand for agent $k$, is the probability that $i$ and $k$ are not direct friends. The second term is the probability of the complement of $k$ not being a friend of any of $i$'s friends, which is exactly the event that $k$ is indeed a friend of a friend of $i$. Notice that the realization of the different links is independent across links, and thus the expectation is just this product.

\paragraph{Proof of Theorem \ref{thm:existence}} Searching for a symmetric equilibrium with positive intensities, assume that all other agents except agent $i$ choose $x\in(0,1]$ as their interaction intensity. Then, agent $i$'s optimization problem is given by:
$$\max_{x_i\in[0,1]}v_1\sum_{k \neq i}p(x_{i},x)+v_2\sum_{k\neq i}\left(1-p(x_{i},x)\right)\left(1-\prod_{l\neq i,k}(1-p(x_{i},x)p(x,x))\right)-\frac{c}{2}\left(\sum_{k\neq i}p(x_{i},x)\right)^2$$

$x_i$ appears in this maximization problem only as an argument of the function $p(\cdot,x)$. Since the function $p(\cdot,x)$ is strictly increasing in its argument, the FOC for the original problem hold if and only if it holds for the problem where we view player $i$ as choosing $p(x_i,x)$, taking $p(x,x)$ as fixed.

Thus, the equilibrium linking probability $p(x,x)=p^*$, if it is internal, must satisfy the following FOC:

$$\frac{\partial}{\partial p}\left. \left(v_1\sum_{k \neq i}p+v_2\sum_{k\neq i}\left(1-p\right)\left(1-\prod_{l\neq i,k}(1-p\cdot p(x,x))\right)-\frac{c}{2}\left(\sum_{k\neq i}p\right)^2\right)\right|_{p(x,x)=p}=0$$

This FOC, after rearrangement, is given by:
\begin{equation} c=\frac{(v_1-v_2)+v_2(1-p)^{n-2}(1+p)^{n-3}(1+(n-1)p)}{(n-1)p} \label{eq:homoFOC} \end{equation}
Since $p(0,0)=0$, and $p(x,y)$ is continuous, the LHS as a function of $p$ tends to positive infinity at 0 (from above), and has derivative:
$$-\frac{(v_1-v_2)+(1-p)^{n-3}(1+p)^{n-4}\left(1+p+(3n-7)p^2+(n-1)(2n-5)p^3\right)v_2}{(n-1)p^2}<0$$
which is strictly negative for $p\in(0,1]$. Thus, if there exists $p^*\in(0,1]$ which solves the FOC, it is the unique symmetric equilibrium intensity. Otherwise, the investment level $p^*=1$ for all agents constitutes a unique symmetric equilibrium. \eproof

\paragraph{Proof of Proposition \ref{prop:compstat}} Implicitly differentiating $p^*$ with respect to $c$ using the FOC (\ref{eq:homoFOC}) yields:
$$\frac{\partial p^*}{\partial c} = -\frac{(n-1)(p^*)^2}{(v_1-v_2)+(1-p^*)^{n-3}(1+p^*)^{n-4}\left(1+p^*+(3n-7)(p^*)^2+(n-1)(2n-5)(p^*)^3\right)v_2}$$
which is strictly negative.

For the comparative static with respect to $v_1$, implicitly differentiating the FOC (\ref{eq:homoFOC}) yields:
$$\frac{\partial p^*}{\partial v_1}=\frac{p^*}{ (v_1-v_2)+\left(1-p^*\right)^{n-3}(1+p^*)^{n-4} \left(1+p^*+(3 n-7) (p^*)^2+(n-1) (2 n-5) (p^*)^3\right) v_2}$$
which is strictly positive.

For the comparative statics with respect to $v_2$, first we again implicitly differentiate $x^*$ with respect to $v_2$:
$$\frac{\partial p^*}{\partial v_2}=\frac{p^*\left((1-p^*)^{n-2}(1+p^*)^{n-3}(1+(n-1)p^*)-1\right)}{ (v_1-v_2)+\left(1-p^*\right)^{n-3}(1+p^*)^{n-4} \left(1+p^*+(3 n-7) (p^*)^2+(n-1) (2 n-5) (p^*)^3\right) v_2}$$
since the denominator is always positive, the sign of the entire derivative is determined by the sign of:
\begin{equation}\left(\left(1-p^*\right)^{n-2} \left(1+p^*\right)^{n-3} \left(1+(n-1)p^*\right)-1\right)\label{eq:compstat}\end{equation}
This expression, as a function of $p^*$, is $0$ at 0, converges to $-1$ from above at 1, and has derivative:
$$-(n-2) (1-p^*)^{n-3} (1+p^*)^{n-4} (2(n-1)(p^*)^2+3p^*-1)$$ 
which is positive just right of 0, and becomes negative before one, changing sign once. Thus, (\ref{eq:compstat}) is first negative and then positive, changing signs only once. This completes the proof. \eproof

\paragraph{Proof of Theorem \ref{thm:maintheorem}} 

Consider the FOC in (\ref{eq:homoFOC}), which we rewrite as:
\begin{equation}
c=\frac{(v_1-v_2)+v_2(1-p)\left(1-p^2\right)^{n-3}(1+(n-1)p)}{(n-1)p} \label{eq:homoFOC2}
\end{equation}

. We first claim that it must hold in the symmetric equilibrium for $n$ large enough. To see this, by the proof Theorem \ref{thm:existence} we have that if the FOC does not hold in equilibrium, then the symmetric equilibrium is the one where all linking probabilities are 1, and hence for all $j$, $x_j=1$. $i$'s marginal utility (in $p(x_i,1)$) at those equilibrium intensity levels, when $x_j=1$ for all $j,k\neq i$ is:
$$\frac{v_1-v_2}{(n-1)}-c$$
which becomes negative for large enough $n$. Thus, for large enough $n$ this could not be the equilibrium, and the symmetric equilibrium must satisfy the FOC.

We now claim that $p^*\to 0$ as $n\to\infty$. Assume otherwise, then there exists a sequence $n_k\to\infty$ with $p^*(n_k)>\epsilon$ for some $\epsilon>0$. Along this subsequence, the denominator of the RHS of (\ref{eq:homoFOC2}) is going to infinity. The numerator of the RHS of (\ref{eq:homoFOC2}) however goes to $v_1-v_2$. Thus, the RHS converges along this subsequence to $0$, yielding a contradiction, as the LHS is fixed at $c$ for any $n$.

Using this, we consider the following cases:
\begin{enumerate}
\item $\limsup_{n\to\infty}n^{\frac{1}{2}}p^*(n)=0$.

This implies that $$\lim_{n\to\infty}\left(1-(p^*)^2\right)^{n-3}=1$$
and thus when $n$ goes to $\infty$ the dominant term in the numerator of the RHS of (\ref{eq:homoFOC2}) is $v_1+v_2(n-1)p^*$, which implies that under this assumption (\ref{eq:homoFOC2}) yields:
\begin{equation}c=\lim_{n\to\infty}\frac{v_1+v_2(n-1)p^*(n)}{(n-1)p^*(n)}\label{eq:FOC3}\end{equation}

If $\liminf p^*(n)n=0$, and this limit is realized along some subsequence $n_k$, then looking at (\ref{eq:FOC3}) along this subsequence yields $c=v_1/0$, which is a contradiction. If $\limsup p^*(n)n=\infty$, and this limit is achieved along some subsequence $n_k$, then (\ref{eq:FOC3}) along this subsequence yields $c=v_2$, which does not hold for any of our cases.

This leaves the case where all the partial limits of $p^*(n)n$ are strictly positive, finite numbers. Let $d$ be such a partial limit, which is achieved along a subsequence $n_k$. Taking the limit of (\ref{eq:FOC3}) along this subsequence yields:
$$c=\frac{v_1+v_2d}{d}$$ which gives $$d = \frac{v_1}{c-v_2}$$
Since the partial limit is pinned down by this equation, it must be that it is the true limit of $p^*(n)n$. Additionally, for this limit to make sense we must have that $c>v_2$.

Summing up, if $\limsup_{n\to\infty}n^{\frac{1}{2}}p^*(n)=0$, then if $v\neq c_2$, we must have that $v_2<c$ and that $\lim_{n\to\infty}p^(n)n =\frac{v_1}{c-v_2}$.

\item $\liminf_{n\to\infty}n^{\frac{1}{4}}p^*(n)=\infty$.

Under this assumption we have that: $$\lim_{n\to\infty}\left(1-(p^*)^2\right)^{n-3}=0$$ and thus the in the limit, the dominant term in the numerator of the RHS of (\ref{eq:homoFOC2}) is $v_1-v_2$, while the denominator converges to infinity. This implies that taking $n\to\infty$ in (\ref{eq:homoFOC2}) yields  $c=0$, a contradiction.

\item Every partial limit of $p^*(n)n^{\frac{1}{2}}$ is a finite, positive number. 

In this case, take some subsequence $n_k$ such that along it we have $p^*(n)n^{\frac{1}{2}}\to d$, for some positive number $d$. Taking a partial limit of (\ref{eq:homoFOC2}) along this subsequence yields:

$$c= \frac{v_2de^{-d^2}}{d}=v_2e^{-d^2}$$

Solving for $d$, we have that:
$$d=\log\left(\frac{v_2}{c}\right)^{\frac{1}{2}}$$
This implies that there is unique possible partial limit for $p^*(n)n^{\frac{1}{2}}$, and thus this sequence converges to $\log\left(\frac{v_2}{c}\right)^{\frac{1}{2}}$. For this to be well defined, we must have $v_2>c$.
\end{enumerate}
Since these three cases exhaust all the possibilities, this completes the proof.
 \eproof

\paragraph{Proof of Proposition \ref{prop:properties}} These results are standard results in random graph theory. See \cite{JacksonBook}, Theorem 4.1 for the results on connectedness and sections 4.2.4, 4.2.5 for the results on the size and existence of the giant component.

For the diameter of the connected network in the high-intensity regime, Corollary 10.12(i) in \cite{Bollobas} gives the result directly. \eproof

\paragraph{Proof of Theorem \ref{thm:efficiency}} The proof follows the same arguments, word-for-word, as the proof of Theorem \ref{thm:maintheorem} applied to the homogeneous agent case, but for the FOC of the social planner instead of that of the individual agent. This FOC is derived by differentiating the agents' utility function with respect to $p$, given that \emph{all} realization probabilities are $p$, and is explicitly given by:
$$c= \frac{v_1-v_2+\left(1-p\right)^{n-2} \left(1+p\right)^{n-3} \left(1+(2 n-3) p\right) v_2}{(n-1) p}.$$ The claim follows. \eproof

\paragraph{Proof of Theorem \ref{thm:hetroexistence}} The proof follows the standard Kakutani fixed-point approach to showing the existence of a symmetric equilibrium when utilities are concave and the choice set of each agent is convex. The only delicate point is that the strategy profile where all agents choose intensity $0$ is an equilibrium, and so one must show that there is an equilibrium in addition to this. This is done by restricting agents to play intensities strictly greater than $\epsilon$ and showing the constraint does not bind. The details will be completed in the final draft. \eproof

\paragraph{Proof of Theorem \ref{thm:hetro}}
Let us first derive the FOC for agent $i$ with private cost coefficient $c_h$. Denote by $x_h^*$ the equilibrium strategy that an agent with cost coefficient $c_h$ uses in equilibrium. Let us denote by $\left\{X_k\right\}_{k\neq i}$ independent random variables which take on value $x_h^*$ with a probability $\textbf{p}(c_h)$. In other words, $X_k$ expresses the intensity of agent $k$ from the perspective of agent $i$, in equilibrium. Agent $i$'s maximizes:

\begin{align*} & v_1x\sum_{k \neq i}\Ex\left[X_k\right]+v_2\sum_{k\neq i}\Ex\left[(1-xX_k)\left(1-\prod_{l\neq i,k}\left(1-xX_l^2X_k\right)\right)\right]-\frac{c_h}{2}\Ex\left[\left(x\sum_{k\neq i}X_k\right)^2\right]
\\ = &(n-1)v_1x\Ex\left[X_1\right]+v_2(n-1)\Ex\left[(1-xX_1)\left(1-\prod_{l\neq i,1}\left(1-xX_l^2X_1\right)\right)\right]-\frac{c_hx^2}{2}\Ex\left[\left(\sum_{k\neq i}X_k\right)^2\right]
\\ = &(n-1)v_1x\Ex\left[X_1\right]+v_2(n-1)\Ex\left[(1-xX_1)\left(1-\left(1-x\Ex\left[X_1^2\right]X_1\right)^{n-2}\right)\right]-\frac{c_hx^2}{2}\Ex\left[\left(\sum_{k\neq i}X_k\right)^2\right]
\end{align*}
by the law of iterated expectation, and the fact that the $X_k$'s are i.i.d. random variables. When taking FOCs, we can legitimately take derivative under the expectation sign, as the distribution over costs is finite. This yields:
\begin{align*} c_h &= \left[ x^*_h\left(\Ex\left[X_1^2\right]+(n-2) \Ex\left[X_1\right]^2\right) \right]^{-1} \Big[(v_1-v_2) \Ex\left[X_1\right] \\ &+v_2\Ex\left[X_1\left(1-x^*_i\Ex\left[X_1^2\right]X_1\right)^{n-2}+(1-x^*_hX_1)(n-2)X_1\Ex\left[X_1^2\right]\left(1-x^*_h\Ex\left[X_1^2\right]X_1\right)^{n-3}\right] \Big]\end{align*}
which can be rewritten as:
\begin{equation}c_h=\frac{(v_1-v_2) \Ex\left[X_1\right]+v_2\Ex\left[X_1\left(1-x^*_h\Ex\left[X_1^2\right]X_1\right)^{n-3}\left(1+(n-2)\Ex\left[X_1^2\right]-(n-1)x^*_hX_1\Ex\left[X_1^2\right]\right)\right]}{x^*_h\left(\Ex\left[X_1^2\right]+(n-2) \Ex\left[X_1\right]^2\right)}\label{hetroFOC}\end{equation}

We first claim that the equilibrium $x_h^*$ goes to 0 with $n$, for all possible costs $c_h\in \mathcal{C}$. In particular this implies that from some $n$ onwards the FOC holds with equality for $x_h^*$ for all $h$. Assume otherwise. This means that the FOC holds with the LHS$\leq$RHS for some $c_h$ and a sequence $n_k\to\infty$, such that for any $n_k$ it holds that $x_h^*>\epsilon>0$. This implies a contradiction, since in (\ref{hetroFOC}) the denominator of the FOC of the RHS grows to infinity:
$$\lim_{k\to\infty}x^*_h\left(\Ex\left[X_1^2\right]+(n_k-2) \Ex\left[X_1\right]^2\right)\geq \lim_{k\to\infty}\epsilon^3p_h(n_k-2)=\infty$$
while the numerator is bounded:
\begin{align*}&\lim_{k\to\infty}(v_1-v_2) \Ex\left[X_1\right]+v_2\Ex\left[X_1\left(1-x^*_h\Ex\left[X_1^2\right]X_1\right)^{n-3}\left(1+(n-2)\Ex\left[X_1^2\right]-(n-1)x^*_hX_1\Ex\left[X_1^2\right]\right)\right] \\ &\leq\lim_{k\to\infty}(v_1-v_2) \Ex\left[X_1\right]+v_2\Ex\left[X_1\left(1-\epsilon\right)^{n-3}\left(1+(n-2)\Ex\left[X_1^2\right]-(n-1)x^*_hX_1\Ex\left[X_1^2\right]\right)\right] \\ &\leq\lim_{k\to\infty}(v_1-v_2) \Ex\left[X_1\right]<\infty\end{align*}

Assume next that $\displaystyle\limsup_{n\to\infty} \max_{1\leq h\leq m}n^{1/4}x_h^*=0$. This implies that for any $h$:
$$\lim_{k\to\infty}\left(1-x^*_h\Ex\left[X_1^2\right]X_1\right)^{n_k-3} =1$$
and thus the dominant term in the numerator of the RHS of (\ref{hetroFOC}) is $v_1\Ex\left[X_1\right]+v_2(n-2)\Ex\left[X_1\right]\Ex\left[X_1^2\right]$, while the dominant term in the denominator is $x_h(n-2)\Ex\left[X_1\right]^2$, giving us that:
\begin{equation}c_h = \lim_{n\to\infty}\frac{v_1+v_2(n-2)\Ex\left[X_1^2\right]}{x_h^*(n-2)\Ex\left[X_1\right]}\label{eq:halfcase}\end{equation}
which also implies that for any $h_1,h_2$:
$$\frac{c_{h_1}}{c_{h_2}}=\lim_{n\to\infty}\frac{x_{h_2}^*}{x_{h_1}^*}$$
so the equilibrium intensities converge together.

Consider three cases:

\begin{enumerate}
	\item If $\displaystyle\liminf_{n\to\infty}x_hn^{1/2}=0\;$ along a subsequence $n_k$, then looking at (\ref{eq:halfcase}) along this subsequence yields $c_h=v_1/0$, which is a contradiction.
	\item If $\displaystyle\limsup_{n\to\infty}x_hn^{1/2}=\infty\;$ along a subsequence $n_k$, then looking at (\ref{eq:halfcase}) along this subsequence yields:
	$$c_h=\lim_{k\to\infty}\frac{v_2\Ex\left[X_1^2\right]}{x_h^*\Ex\left[X_1\right]}$$
	Inverting this formula, and writing in terms of sociability instead of cost, yields:
	$$s_h =\lim_{k\to\infty}\frac{x_h^*\Ex\left[X_1\right]}{v_2\Ex\left[X_1^2\right]} \qquad s_h^2 =\lim_{k\to\infty}\frac{(x_h^*)^2\Ex\left[X_1\right]^2}{v_2^2\Ex\left[X_1^2\right]^2}$$
	
	Taking expectation with respect to $\textbf{p}$ of both terms and dividing the first by the second, yields:
	$$v_2=\frac{\Ex\left[S\right]}{\Ex\left[S^2\right]}$$
	which does not hold for any of our cases.
	\item Assume that for some $h$ some partial limit along a subsequence $n_k$ satisfies $\lim_{k\to\infty}x_h^*n^{1/2}=d_h>0$. Because the $x_h^*$ converge together, we have that for any $h'$ there 
	exists some $d_{h'}$ such that $\lim_{k\to\infty}x^*_{h'}n^{1/2}=d_{h'}>0$. (\ref{eq:halfcase}) thus becomes, with slight abuse of notation:
	\begin{equation}c_h=\frac{v_1+v_2\Ex\left[d^2\right]}{d_h\Ex\left[d\right]}\label{eq:halfcase2}\end{equation}
	which also implies that for any $h,h'$ it holds that $d_{h'}=d_hc_h/c_{h'}$. Plugging the expressions for the other $h'$ back into (\ref{eq:halfcase2}) yields:
	$$c_h=\frac{v_1+v_2d_h^2c_h^2Ex\left[S^2\right]}{d_h^2c_h\Ex\left[S\right]}$$
	which is solved by:
	$$d_h=s_h\left(\frac{v_1}{\Ex\left[S\right]-v_2\Ex\left[S^2\right]}\right)^{\frac{1}{2}}$$ 
	Since the limit is pinned down, it is the true limit of the sequence $x_h^*n^{1/2}$. If $v2>\Ex\left[S\right]/\Ex\left[S^2\right]$, then we have a contradiction.
\end{enumerate}

The above shows that if $\displaystyle\limsup_{n\to\infty} \max_{1\leq h\leq m}n^{1/4}x_h^*=0$, then $v_2<\Ex\left[S\right]/\Ex\left[S^2\right]$ and the limits in part (1) of the theorem hold.

Next, assume that $\displaystyle\liminf_{n\to\infty} \min_{1\leq h\leq m}n^{1/4}x_h^*=\infty$, along some subsequence $n_k$. This implies that:
$$\lim_{k\to\infty}\left(1-x^*_h\Ex\left[X_1^2\right]X_1\right)^{n_k-3} =0$$
and thus the dominant term in the numerator of the RHS of (\ref{hetroFOC}) is $(v_1-v_2)\Ex\left[X_1\right]$, while the dominant term in the denominator is $x_h(n_k-2)\Ex\left[X_1\right]^2$, giving us that:
$$c_h=\lim_{k\to\infty}\frac{v_1-v_2}{(n_k)x_h\Ex\left[X_1\right]}=0$$
by the assumption that $\lim_{k}\max_{1\leq h\leq m}n^{1/4}x_h^*=0$. This is a contradiction.

Thus, the only remaining case is that for every $h$ and every subsequence $n_k\to\infty$ we have that $\lim_k x_h^*n^{1/4}$ converges to a finite positive number, $d_h$. Assuming this is the case, the FOC in (\ref{hetroFOC}):
\begin{equation}c_h=\frac{v_2\Ex\left[d^2\right]\Ex\left[de^{-d_hd\Ex\left[d^2\right]}\right]}{d_h\Ex\left[d\right]^2}\label{eq:hetrolimit}\end{equation}
where the expectations are taken over $d$ as a random variable which assumes value $d_h$ with probability $\textbf{p}(c_h)$. Inverting this equation and taking expectation with respect to $\textbf{p}$ yields:
$$\Ex\left[S\right]=\Ex\left[\frac{d'\Ex\left[d\right]^2}{v_2\Ex\left[d^2\right]\Ex\left[de^{-d'd\Ex\left[d^2\right]}\right]}\right]\qquad 
\Ex\left[S^2\right]=\Ex\left[\frac{(d')^2\Ex\left[d\right]^4}{v_2^2\Ex\left[d^2\right]^2\Ex\left[de^{-d'd\Ex\left[d^2\right]}\right]^2}\right]$$ 
where the outer expectation is taken over $d'$, a random variable which is distributed identically to $d$. Taking the ratio of these two expressions yields:
\begin{equation}\frac{\Ex\left[S\right]}{\Ex\left[S^2\right]}=v_2\frac{\Ex\left[\frac{d'\Ex\left[d\right]^2}{\Ex\left[d^2\right]\Ex\left[de^{-d'd\Ex\left[d^2\right]}\right]}\right]}
{\Ex\left[\frac{(d')^2\Ex\left[d\right]^4}{\Ex\left[d^2\right]^2\Ex\left[de^{-d'd\Ex\left[d^2\right]}\right]^2}\right]}\label{eq:lemmacause}\end{equation}

We will make use of the following lemma.

\begin{lemma} \label{hetrolemma}
$$\Ex\left[\frac{d'\Ex\left[d\right]^2}{\Ex\left[d^2\right]\Ex\left[de^{-d'd\Ex\left[d^2\right]}\right]}\right]
<\Ex\left[\frac{(d')^2\Ex\left[d\right]^4}{\Ex\left[d^2\right]^2\Ex\left[de^{-d'd\Ex\left[d^2\right]}\right]^2}\right].$$
\end{lemma}

By Lemma \ref{hetrolemma} and (\ref{eq:lemmacause}) we have that:
$$\frac{\Ex\left[S\right]}{\Ex\left[S^2\right]}<v_2$$ and thus when $\lim_k x_h^*n^{1/4}$ converges to a positive number, we must be in the high-intensity regime. This completes the proof. \eproof

\paragraph{Proof of Lemma \ref{hetrolemma}} It holds that:
$$\Ex\left[\frac{d'\Ex\left[d\right]^2}{\Ex\left[d^2\right]\Ex\left[de^{-d'd\Ex\left[d^2\right]}\right]}\right]
-\Ex\left[\frac{(d')^2\Ex\left[d\right]^4}{\Ex\left[d^2\right]^2\Ex\left[de^{-d'd\Ex\left[d^2\right]}\right]^2}\right]<0$$
if and only if:
$$\Ex\left[\frac{1}{\Ex\left[de^{-d'd\Ex\left[d^2\right]}\right]}\left(d'
-\frac{(d')^2\Ex\left[d\right]^2}{\Ex\left[d^2\right]\Ex\left[de^{-d'd\Ex\left[d^2\right]}\right]}\right)\right]<0$$
Since $e^{-dd'\Ex\left[d^2\right]}\leq 1$ as $d,d'$ are strictly positive random variables, it holds that:
$$\Ex\left[\frac{1}{\Ex\left[de^{-d'd\Ex\left[d^2\right]}\right]}\left(d'
-\frac{(d')^2\Ex\left[d\right]^2}{\Ex\left[d^2\right]\Ex\left[de^{-d'd\Ex\left[d^2\right]}\right]}\right)\right]< \Ex\left[\frac{1}{\Ex\left[de^{-d'd\Ex\left[d^2\right]}\right]}\left(d'
-\frac{(d')^2\Ex\left[d\right]^2}{\Ex\left[d^2\right]\Ex\left[d\right]}\right)\right] $$
\begin{equation}= \Ex\left[\frac{1}{\Ex\left[de^{-d'd\Ex\left[d^2\right]}\right]}\left(d'-\frac{(d')^2\Ex\left[d\right]}{\Ex\left[d^2\right]}\right)\right]\label{eq:punchline}\end{equation}

$\left(d'-\frac{(d')^2\Ex\left[d\right]}{\Ex\left[d^2\right]}\right)$ is a random variable with expectation $0$, which is positive for values of $d'$ satisifying $0<d'< \frac{\Ex\left[d^2\right]}{\Ex\left[d\right]}$, and negative for higher values. $1/\Ex\left[de^{-d'd\Ex\left[d^2\right]}\right]$ is a positive function in $d'$ which is strictly increasing. Thus, (\ref{eq:punchline}) must be strictly negative, which completes the proof.

\paragraph{Proof of Proposition \ref{prop:unistable}} The fact that $\tilde{c} \leq v_1-v_2$ is sufficient for unilateral stability is obvious. Now suppose that $\tilde{c} > v_1 - v_2$. We will show that the probability that some agent wants to sever a link is bounded away from $0$ for all $n$. It will suffice to this end to show that the probability of an \emph{isolated triangle} occurring in the network is bounded away from $0$. An isolated triangle is a triple of agents $\{i,j,k\}$ with links $ij$, $jk$, and $ik$, and no links to anyone else. It is clear that if $\tilde{c} > v_2 - v_1$, then any agent in an isolated triangle wants to sever a link. The probability that three given vertices $\{i,j,k\}$ form an isolated triangle is at least \begin{equation} \underline{p}^3(1-\overline{p})^{3(n-2)} \label{eqn:triangleprob} \end{equation} where $\underline{p}$ is the minimum probability of a link between two vertices and $\overline{p}$ is the maximum probability of such a link. In (\ref{eqn:triangleprob}), the first factor is a lower bound on the probability that the three requisite links exist, and the second term is a lower bound on the probability that the agents are not linked to anyone else. Now, there are $\binom{n}{3}$ possible triples of agents, so (by linearity of expectation) the expected number of isolated triangles is at least $\binom{n}{3} \underline{p}^3((1-\overline{p})^{3(n-2)}$. Since both $\underline{p}$ and $\overline{p}$ behave as $n^{-1}$ by Theorem \ref{thm:hetro}, this expectation is bounded away from $0$, as $n\to \infty$.  \eproof


\newpage
\bibliography{formation-bibliography}

\newcommand{\noopsort}[1]{} \newcommand{\printfirst}[2]{#1}
  \newcommand{\singleletter}[1]{#1} \newcommand{\switchargs}[2]{#2#1}
\begin{thebibliography}{22}
\newcommand{\enquote}[1]{``#1''}
\expandafter\ifx\csname natexlab\endcsname\relax\def\natexlab#1{#1}\fi

\bibitem[\protect\citeauthoryear{Ambrus, Mobius, and Szeidl}{Ambrus
  et~al.}{2010}]{ams}
\textsc{Ambrus, A., M.~Mobius, and A.~Szeidl} (2010): \enquote{Consumption
  Risk-sharing in Social Networks,} Preprint, available at \newline
  \verb!http://www.econ.berkeley.edu/~szeidl/papers/risksharing.pdf!

\bibitem[\protect\citeauthoryear{Anon.}{Anon.}{2010}]{facebook}
\textsc{Anon.} (2010): \enquote{Facebook: Statistics,}
  \texttt{http://www.facebook.com/press/info.php?statistics}.

\bibitem[\protect\citeauthoryear{Bala and Goyal}{Bala and
  Goyal}{2000}]{BalaGoyal2000}
\textsc{Bala, V. and S.~Goyal} (2000): \enquote{A Noncooperative Model of
  Network Formation,} \emph{Econometrica}, 68, 1181--1229.

\bibitem[\protect\citeauthoryear{Bollob\'{a}s}{Bollob\'{a}s}{2001}]{Bollobas}
\textsc{Bollob\'{a}s, B.} (2001): \emph{Random Graphs, Second Edition},
  Cambridge: Cambridge University Press.

\bibitem[\protect\citeauthoryear{Boyd and Ellison}{Boyd and
  Ellison}{2007}]{boyd:sns}
\textsc{Boyd, D. and N.~B. Ellison} (2007): \enquote{Social Network Sites:
  Definition, History, and Scholarship,} \emph{Journal of Computer-Mediated
  Communication}, 13, 210--230.

\bibitem[\protect\citeauthoryear{Cabrales, Calv\'{o}-Armengol, and
  Zenou}{Cabrales et~al.}{2009}]{conferencesZenou}
\textsc{Cabrales, A., A.~Calv\'{o}-Armengol, and Y.~Zenou} (2009):
  \enquote{Social Interactions and Spillovers,} Preprint, available at
  \verb!http://www.ifn.se/web/yvesz.aspx!

\bibitem[\protect\citeauthoryear{Calv\'{o}-Armengol, Patacchini, and
  Zenou}{Calv\'{o}-Armengol et~al.}{2009}]{CPZ}
\textsc{Calv\'{o}-Armengol, A., E.~Patacchini, and Y.~Zenou} (2009):
  \enquote{Peer Effects and Social Networks in Education,} \emph{Review of
  Economic Studies}, 76, 1239--1267.

\bibitem[\protect\citeauthoryear{Chung and Lu}{Chung and Lu}{2002}]{chunglu}
\textsc{Chung, F. and L.~Lu} (2002): \enquote{The Average Distance in Random
  Graphs with Given Expected Degrees,} \emph{Proceedings of National Academy of
  Science}, 99, 15879--15882.

\bibitem[\protect\citeauthoryear{Chung, Lu, and Vu}{Chung
  et~al.}{2004}]{ChungLuVu}
\textsc{Chung, F., L.~Lu, and V.~Vu} (2004): \enquote{The Spectra of Random
  Graphs with Given Expected Degrees,} \emph{Internet Mathematics}, 1,
  257--275.

\bibitem[\protect\citeauthoryear{Duflo and Saez}{Duflo and Saez}{2003}]{duflo}
\textsc{Duflo, E. and E.~Saez} (2003): \enquote{The Role of Information and
  Social Interactions in Retirement Plan Decisions: Evidence From a Randomized
  Experiment*,} \emph{Quarterly Journal of Economics}, 118, 815--842.

\bibitem[\protect\citeauthoryear{Erd\H{o}s and R\'{e}nyi}{Erd\H{o}s and
  R\'{e}nyi}{1959}]{ErdosRenyi}
\textsc{Erd\H{o}s, P. and A.~R\'{e}nyi} (1959): \enquote{On Random Graphs,}
  \emph{Publicationes Mathematicae Debrecen}, 6, 290--297.

\bibitem[\protect\citeauthoryear{Glaeser, Sacerdote, and Scheinkman}{Glaeser
  et~al.}{1996}]{glaeser}
\textsc{Glaeser, E., B.~Sacerdote, and J.~Scheinkman} (1996): \enquote{Crime
  and Social Interactions,} \emph{Quarterly Journal of Economics}, 111,
  507--–548.

\bibitem[\protect\citeauthoryear{Granovetter}{Granovetter}{2005}]{Granovetter_%
JEP}
\textsc{Granovetter, M.} (2005): \enquote{The Impact of Social Structure on
  Economic Outcomes,} \emph{The Journal of Economic Perspectives}, 19, 33--50.

\bibitem[\protect\citeauthoryear{Hojman and Szeidl}{Hojman and
  Szeidl}{2008}]{szeidl}
\textsc{Hojman, D. and A.~Szeidl} (2008): \enquote{Core and Periphery in
  Networks,} \emph{Journal of Economic Theory}, 139, 295--309.

\bibitem[\protect\citeauthoryear{Jackson}{Jackson}{2005}]{JacksonFormationSurv%
ey}
\textsc{Jackson, M.~O.} (2005): \enquote{A Survey of Models of Network
  Formation: Stability and Efficiency,} in \emph{Group Formation in Economics:
  Networks, Clubs, and Coalitions}, ed. by G.~Demange and M.~Wooders,
  Cambridge: Cambridge University Press.

\bibitem[\protect\citeauthoryear{Jackson}{Jackson}{2008}]{JacksonBook}
---\hspace{-.1pt}---\hspace{-.1pt}--- (2008): \emph{Social and Economic
  Networks}, Princeton, N.J.: Princeton University Press.

\bibitem[\protect\citeauthoryear{Jackson and Wolinsky}{Jackson and
  Wolinsky}{1996}]{jackson-wolinsky-96}
\textsc{Jackson, M.~O. and A.~Wolinsky} (1996): \enquote{A Strategic Model of
  Social and Economic Networks,} \emph{Journal of Economic Theory}, 71, 44--74.

\bibitem[\protect\citeauthoryear{K\"{o}nig, Tessone, and Zenou}{K\"{o}nig
  et~al.}{2009}]{formationZenou}
\textsc{K\"{o}nig, M.~D., C.~J. Tessone, and Y.~Zenou} (2009): \enquote{A
  Dynamic Model of Network Formation with Strategic Interactions,} Preprint,
  available at \verb!http://www.ifn.se/web/yvesz.aspx!

\bibitem[\protect\citeauthoryear{Lombardi}{Lombardi}{2007}]{NYTimesTransformed}
\textsc{Lombardi, K.~S.} (2007): \enquote{Make New Friends Online, and You
  Won't Start College Friendless,} \emph{New York Times}, {M}arch 21.

\bibitem[\protect\citeauthoryear{Myerson}{Myerson}{1991}]{myerson}
\textsc{Myerson, R.~B.} (1991): \emph{Game Theory: Analysis of Conflict},
  Cambridge, MA: Harvard University Press.

\bibitem[\protect\citeauthoryear{Newman}{Newman}{2003}]{NewmanSurvey}
\textsc{Newman, M. E.~J.} (2003): \enquote{The Structure and Function of
  Complex Networks,} \emph{SIAM Review}, 45, 167--256.

\bibitem[\protect\citeauthoryear{Topa}{Topa}{2001}]{topa}
\textsc{Topa, G.} (2001): \enquote{Social Interactions, Local Spillovers and
  Unemployment,} \emph{Review of Economic Studies}, 68, 261--295.

\end{thebibliography}

\end{document}